\documentstyle[aps,pra,amssymb,psfig]{revtex}

\begin{document}

\draft
\twocolumn[\hsize\textwidth\columnwidth\hsize\csname @twocolumnfalse\endcsname%
\title{Svortices and the fundamental modes of the ``snake instability'':
Possibility of observation in the gaseous Bose--Einstein Condensate}
\author{Joachim Brand and William P. Reinhardt}
\address{Department of Chemistry, University of Washington, Seattle,
WA 98195-1700, USA} 
\date{\today}

\maketitle

\begin{abstract}
The connection between quantized vortices and dark solitons in a long
and thin, waveguide-like trap geometry is explored in the framework of
the non-linear Schr\"odinger equation. Variation of the transverse
confinement leads from the quasi-1D regime where solitons are stable
to 2D (or 3D) confinement where soliton stripes are subject to a
transverse modulational instability known as the ``snake
instability''. We present numerical evidence of a regime of
intermediate confinement where solitons decay into single, deformed
vortices with solitonic properties, also called svortices, rather than
vortex pairs as associated with the ``snake'' metaphor. Further relaxing the
transverse confinement
leads to production of 2
and then 3 vortices, which correlates perfectly with a
Bogoliubov--de Gennes stability analysis. The decay of a
stationary dark soliton (or, planar node) into a single svortex is
predicted to be experimentally observable in a 3D harmonically
confined dilute gas Bose--Einstein condensate.
\end{abstract}
\pacs{03.75.-b, 05.45.Yv, 42.65.Tg}
]
\narrowtext


Solitons and quantized vortices are fundamental excitations of
non-linear media. Quantized vortices, often regarded as an indicator
for superfluidity, are topological defects in (2+1)- or
(3+1)-dimensional fluids. Dark solitons, in their
purest form are solitary, nondispersive density-notch solutions to
(1+1)-dimensional, non-linear wave equations with extraordinary
stability properties. It has been known, however, for many years that
solitonic wave fronts (also called band solitons or soliton stripes)
in 2- or 3-dimensional media are unstable
\cite{zakharov73,jones86,kuznetsov88,law93,kivshar00}. The metaphor of a
``snake'' instability (SI) has been introduced in this context by Zakharov
\cite{zakharov73} in order to refer to the antisymmetric modulation
(bending) of the solitonic wave-front caused by long-wavelength
perturbations \cite{kuznetsov88}. Later it has been predicted by
numerical studies of the time evolution that the SI eventually leads
to the formation of arrays of vortices with alternating charge
\cite{jones86,law93}. 
The first experimental evidence of the SI and subsequent formation of
vortices was observed in non-linear optics
\cite{mamaev96,tikhonenko96}.

\begin{figure}
\begin{center}
\psfig{figure=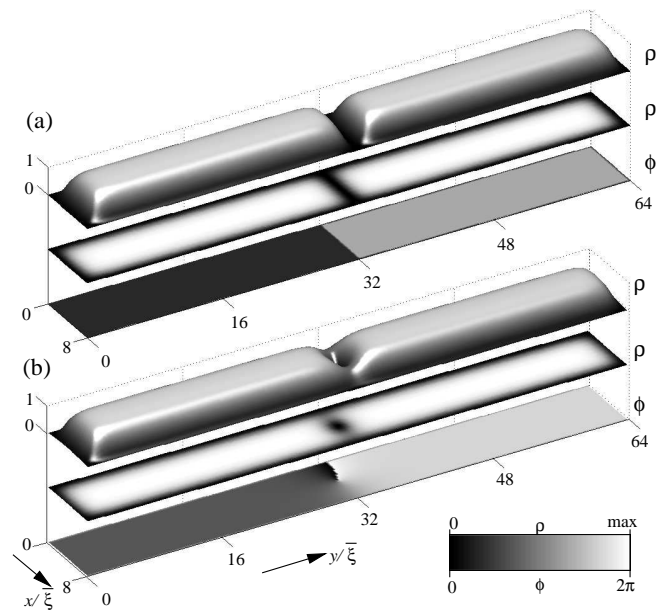}
\end{center}
\caption{Two stationary ``one-defect'' excited states in a
2D rectangular box trap with hard-wall boundary conditions generated
by imaginary-time propagation of the NLSE: (a) stationary soliton (line
node) and (b) stationary svortex state. The complex wavefunction $\psi$
is represented by the density $\rho = |\psi|^2$ and the phase $\phi =
\arg(\psi)$. Each subplot shows a surface plot and gray-scale coded
plots of the density and the phase modulo $2\pi$.}
\label{svortwf}
\end{figure}

More recently, dark solitons have been observed in trapped dilute-gas BECs
\cite{solitonexp} and the decay of a stationary
soliton into closed loops of vortex filaments, much resembling smoke
rings, has been observed in a spherical harmonic trap
\cite{Anderson2001a}.
Stationary dark solitons, like the example shown in
Fig.~\ref{svortwf}(a), are nodes (nodal lines or planes in 2D or 3D,
respectively) in the wavefunction as opposed to traveling solitons,
which are also referred to as gray solitons.  Theoretically, the
stability of stationary solitons in harmonically trapped BECs had been
investigated before by Muryshev {\it et al.}\ \cite{Muryshev1999a} and
Feder {\it et al.}\ \cite{Feder2000a}, based on a linear stability
analysis using the Bogoliubov--de Gennes (BdG) equations.  While both
papers identify a regime of stability for stationary solitons in
elongated traps at low density as expected from earlier work
\cite{kuznetsov88}, it was conjectured in Ref.~\cite{Muryshev1999a}
that the mechanism of instability at increasing density was vortex
pair production in analogy to the SI. Feder {\it et al.}\
\cite{Feder2000a} refined and partially corrected the results of
Ref.~\cite{Muryshev1999a} and predicted the later experimentally
observed vortex ring formation \cite{Anderson2001a}. The mechanism of
decay at the onset of instability, however, has not been fully
revealed so far.

In this letter we study the modes of instability of a stationary
soliton as a function of the transverse confinement ${L}_{\rm t}$,
measured in terms of the condensate healing length $\xi$
\cite{Reinhardt1997a,brand01a}.  The onset of instability at ${L}_{\rm
t} \gtrapprox 6 \xi$ is initiated by the emergence of a nontrivial
stationary state [see Fig.~\ref{svortwf}(b)] of {\em lower} energy
than the corresponding stationary soliton. We call this state a
solitonic vortex or ``svortex''. A svortex is a single confined and
deformed vortex with solitonic properties \cite{brand01a}.  For
transverse confinements of $6 \xi \lessapprox {L}_{\rm t} \lessapprox
10 \xi$ the strong coupling of the stationary soliton to the more
stable single svortex is the only decay mechanism available in
contrast to what has been seen and expected in earlier
work\cite{kuznetsov88,Muryshev1999a,Feder2000a,Anderson2001a}. The
svortex therefore presents the smallest possible unit of decay, which
persists in geometries where the transverse confinement is too
tight for vortex ring (in 2D: vortex pair) formation. Under less
restrictive confinement, 2 and then 3 vortex channels open (for
${L}_{\rm t} \gtrapprox 10 \xi$ and ${L}_{\rm t} \gtrapprox 13 \xi$,
respectively). The 1, 2, and 3 vortex instabilities will be seen to
correlate perfectly with a BdG stability analysis.

The essential physics involved reveals itself from studying the
time-dependent Gross-Pitaevskii or nonlinear Schr\"odinger equation
(NLSE), which presents the relevant mean-field theory for a
zero-temperature BEC\cite{Dalfovo1999a} and also applies to non-linear
wave propagation in optics \cite{kivshar98}:
\begin{equation} \label{nlse}
  i \partial_t \psi = [-\nabla^2 + V + g v_{\rm B} |\psi|^2]\psi .
\end{equation}
In the dimensionless Eq.~(\ref{nlse}), the condensate wavefunction
$\psi(${\bf r}$, t)$ satisfies the following normalization condition:
$\int_{v_{\rm B}} |\psi|^2 d${\bf r}$ = 1$, where ${v_{\rm B}}$ is the
volume of a box containing the trapped condensate. The external
trapping potential is given by $V$ and $g$ is the nonlinear coupling
constant. We restrict ourselves to a repulsive nonlinearity (or
defocusing NLSE) $g > 0$. The relevant size scale for nonlinear
structures like solitons\cite{kivshar98} and vortices\cite{fetter01}
is the condensate healing length $\xi = 1/\sqrt{g {v_{\rm B}}
|\psi|^2} \tilde{\xi}$, where $\tilde{\xi} = 8 \pi a N /(g {v_{{\rm
B}}})$ is unit of length used in Eq.~(\ref{nlse}) for a BEC with $N$
particles and an $s$-wave scattering length $a$. Note that for fairly
uniformly distributed condensates, the healing length is given by
$\overline{\xi} =1/\sqrt{g} \tilde{\xi}$. The application of NLSE
solutions of type of Fig.~\ref{svortwf} has been fully confirmed
empirically by the experiments of
Refs.~\cite{solitonexp,Anderson2001a}. In tightly confined BECs, the
current mean-field theory is justified as long as the transverse
dimensions are greater than $\xi$ and $\xi \gg a$ is satisfied (see
\cite{petrov00}).

We initially consider a 2D rectangular geometry with box boundary
conditions. The chosen aspect ratio length/width of 8 simulates a
transversely confined, waveguide-like geometry. The stationary
vortex-like state with a node and phase singularity at the trap center
was found by imaginary-time propagation and confirmed by real-time
propagation of the NLSE \cite{brand01a}. In addition to seeding this
relaxation procedure with a suitable phase profile, we also restricted
the symmetry of the density $|\psi|^2$ to be even in both spacial
directions. A second stationary state (a dark band soliton) was also
generated by imaginary-time propagation with the constraint of odd
symmetry in the longitudinal direction of the trap.  Figure
\ref{svortwf} shows the resulting wavefunctions. The vortex-like
wavefunction of Fig.~\ref{svortwf}(b) is clearly distorted and
affected by the tight traverse confinement of $8 \overline{\xi}$. We
have argued in Ref.~\cite{brand01a} that such a tightly confined
vortex acquires solitonic properties and therefore should be called
solitonic vortex or svortex, further discussed in \cite{tbp}.

\begin{figure}
\begin{center}
\psfig{figure=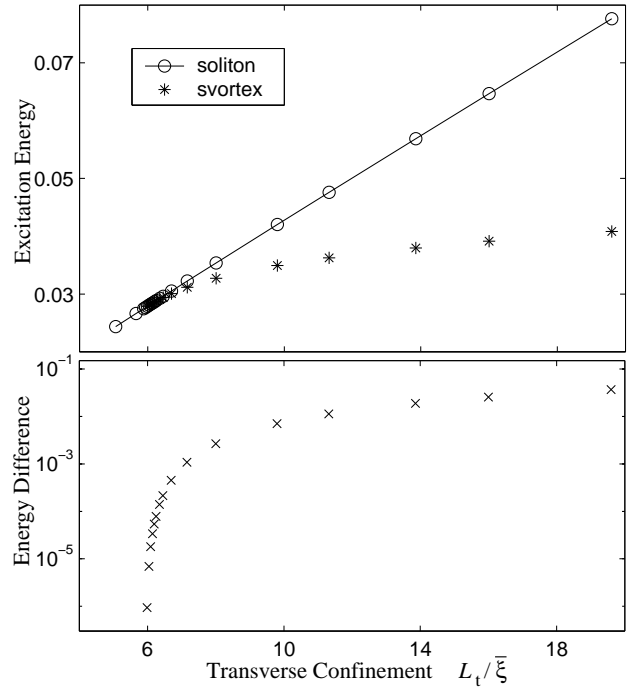}
\end{center}
\caption{Svortex and soliton properties as a function of the
transverse confinement ${L}_{\rm t}$. Part (a) shows the excitation
energies of the stationary soliton and svortex state. The energy
difference is shown on a logarithmic scale in part (b). The
simulations were done in a rectangular box of size $8\tilde{\xi}\times
64\tilde{\xi}$ for different values of the nonlinear coupling constant
$g$, which changes the effective transverse confinement ${L}_{\rm
t}/\overline{\xi} = 8 \sqrt{g}$.
}
\label{energyfiga}
\end{figure}

Figure \ref{energyfiga}(a) shows the excitation energy of the
stationary svortex and soliton state with respect to the ground state
as a function of the transverse confinement ${L}_{\rm t}$ in terms of
$\overline{\xi}$. For a given wavefunction $\psi$, the energy is given
by the formula $E = \int_{v_{\rm B}} (-\psi^* \nabla^2 \psi + g
{v_{\rm B}}/2 |\psi|^4 + V |\psi|^2)$ d{\bf r}. The soliton excitation
energy exhibits a linear dependence on the variation of the length
scale reflecting the localization of the excitation in one and
extension in the other spacial dimension. The svortex energy is always
lower than the soliton energy and grows more slowly with the box size,
reflecting the expected logarithmic behavior for large boxes
\cite{fetter01}. Below a critical confinement corresponding to a box width
of $\approx 6 \overline{\xi}$ we do not find any stationary svortex
solutions but instead the imaginary-time propagation converges to the
soliton solution. The logarithmic plot of the energy difference
between the soliton and the svortex energies shown in
Fig.~\ref{energyfiga}(b) very much indicates a nonanalytic curve
joining or curve crossing. Following the svortex solution from wide
confinement to the critical point, the vortex wavefunction shows an
increasingly deformed density and squeezed phase signature (see
Fig.~\ref{svortwf}) and eventually coincides with the soliton
wavefunction at the critical confinement.

The band solitons and svortex states from Fig.~\ref{energyfiga}(a) are
stationary states. In wide enough confinement, however, band solitons
may exhibit the SI mentioned earlier: Tiny imperfections of stationary
band solitons may lead to a transverse modulation and grow during
real-time propagation at an initially exponential rate. The stability
of stationary solutions of the NLSE can be tested in a linear
stability analysis employing the famous BdG
equations \cite{Bogolubov1947a}, which can be derived from a
linear-response expansion of the time-dependent NLSE
\cite{Edwards1996a}. In the units of Eq.~\ref{nlse} the BdG equations
read:
\begin{eqnarray} \label{bogeqn}
  {\cal{L}} u_j(\bbox{r}) - g v_{\rm B} [\psi(\bbox{r})]^2
  v_j(\bbox{r}) &=& \epsilon_j u_(\bbox{r})j, \\ 
  {\cal{L}} v_j(\bbox{r}) - g v_{\rm B} [\psi^*(\bbox{r})]^2
  u_j(\bbox{r}) &=& -\epsilon_j v_j(\bbox{r}), 
\end{eqnarray}
with ${\cal{L}} = -\nabla^2 + V(\bbox{r}) + 2 g v_{\rm B}
|\psi(\bbox{r})|^2 - \mu,$ and $\mu$ is the chemical potential of the
stationary wavefunction $\psi(\bbox{r},t) = \exp(-i \mu t)
\psi(\bbox{r})$. The solutions of the BdG equation with eigenvalues
$\epsilon_j$ and eigenvectors $(u_j, v_j)$ have the following
interpretation in terms of small-amplitude motion around a stationary
solution of the NLSE \cite{fetter01}: Small positive $\epsilon_j$ at
positive ``norm'' $\eta_j = \int (|u_j|^2 - |v_j|^2) $d{\bf r}
describe small oscillations around the stationary state with
increasing energy. Solutions with negative eigenvalue $\epsilon_j$ and
positive $\eta_j$ are called anomalous modes. They indicate a
continuous transformation of the stationary state to a state of lower
energy. Anomalous modes exist for the trapped vortex as well as for
dark solitons in 1D and merely express the thermodynamic instability
of these excitations. Complex or purely imaginary eigenvalues
$\epsilon_j$, however, indicate a dynamical instability. They further
imply $\eta_j=0$.

\begin{figure}
\begin{center}
\psfig{figure=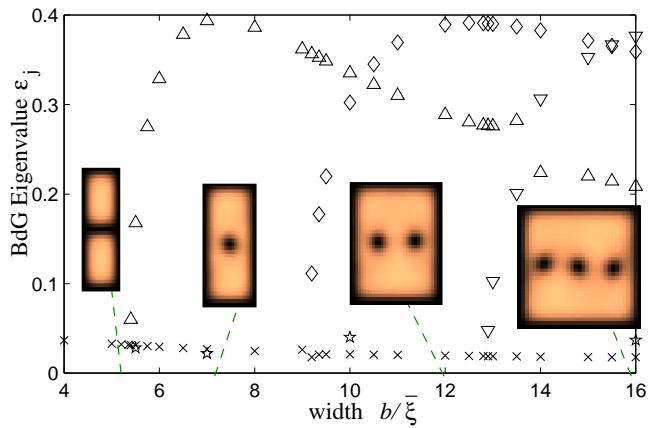}
\vspace{-5mm}
\end{center}
\caption{Eigenvalues in the BdG spectrum of the stationary soliton in
a 2D box as a function of the box width at constant average
density. The insets relating to box widths of $b= 7\overline{\xi}$,
$12\overline{\xi}$, and $16\overline{\xi}$ show density plots of
transient patters in the decay of the perturbed soliton state [21]
after real-time propagation for $t=26$, 31, and 26 in the
units of Eq.~\ref{nlse}, respectively. The perturbed soliton at
$b=5\overline{\xi}$, on the contrary, shows no appreciable decay after
100 time units. The imaginary modes are marked according to the nature
of the eigenvector $u$ leading to single-vortex ($\triangle$),
double-vortex ($\diamond$) 
or triple-vortex ($\triangledown$) decay. The anomalous mode of the nodal
plane-state ($\times$) and stationary single-svortex state ($\bigstar$) are
also indicated.}
\label{bogspec}
\end{figure}

Figure \ref{bogspec} shows the purely imaginary and anomalous
eigenvalues of the BdG equation for a stationary band soliton in a
rectangular box of dimension $b \times 16 \overline{\xi}$ as a
function of the box width $b \approx {L}_{\rm t}$ at constant
density. For narrow traps with $b \lessapprox 5.5 \overline{\xi}$, we
find one anomalous but no complex eigenvalues, like for 1D
solitons. Additionally, the soliton wavefunction shows no appreciable
decay in real-time propagation seeded with noise (see insets).  Also
collisions of noisy gray solitons show the robust, particle-like
behavior expected from 1D soliton theory \cite{tbp}. For trap widths
$5.5 \overline{\xi}\lessapprox b \lessapprox 9.5 \overline{\xi}$ one
purely imaginary eigenvalue exists in the BdG eigenvalue
spectrum. According to the numerical results, the emergence of this
imaginary eigenvalue coincides with the emergence of the svortex as a
symmetry-breaking stationary state of lower energy than the
corresponding band soliton. Increasing the box width, a second and
eventually a third imaginary eigenvalue appears. The stability of the
stationary soliton was probed using real-time propagation seeded with
a small amount of white noise \cite{noise}. While there is no
appreciable decay in tight confinement, we clearly find that the
soliton instability is associated with the formation of one, two, and
three vortices in the regimes where one, two, and three imaginary
eigenvalues are present as shown in the insets of
Fig.~\ref{bogspec}. The BdG eigenvectors $u_j$, localized within about
one healing length from the nodal line of the soliton, also support
this result \cite{tbp}. The patterns shown in Fig.~\ref{bogspec} are
by no means stationary but rather form transient states followed by
incomplete recurrences of the nodal line and eventual further decay
where vortices move to edge of the trap and vorticity is
destroyed. The complicated dynamical patterns showing a mixture of
decay and strong mode coupling are certainly due to energy
conservation in the NLSE and to the small scale of the trap used in
the simulation where radiated phonons linger. We expect further
stabilization of the vortex patterns in longer traps where energy
released in the decay process can distribute itself over a larger
area. The observed decay patterns vary depending on the exact form of
the initial perturbation by noise. In contrast to the soliton, the
stationary svortex shows an entirely real BdG spectrum with one
anomalous mode also shown in Fig.~\ref{bogspec}. Further, real-time
propagation of perturbed svortices shows no appreciable decay. In this
sense, the svortex is the more stable object than the stationary
soliton.

\begin{figure}
\begin{center}
\psfig{figure=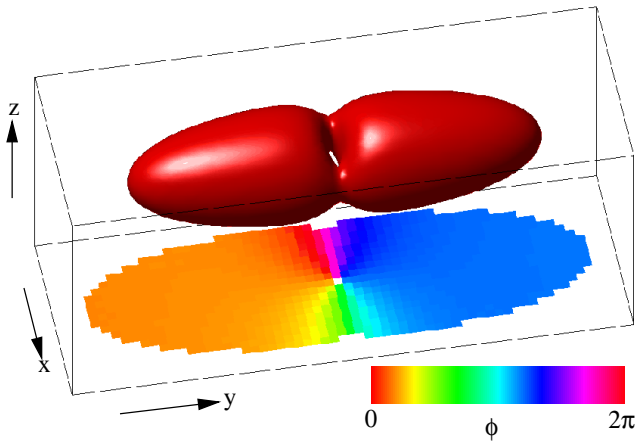}
\end{center}
\caption{\hspace{-3mm}(color). Svortex in a 3D elongated harmonic trap
generated by decay of 
a perturbed stationary soliton state. Shown is the surface of constant
density (at 0.16 of the maximum density) and a color-coded plot of the
phase in the horizontal plane intersecting the trap center. The
transverse confinement ${L}_{\rm t}/\xi \approx 7.7$ was computed
by the maximum value of the line integral $\int_{{\cal
C}}\xi(\mbox{\bf r})^{-1}\, {\rm ds}$ taken along the transverse dimension,
which is more appropriate for measuring the transverse confinement of
inhomogeneous condensates
than the box width [13].
For details of the simulation see
text.}
\label{3dharmonic}
\end{figure}

Finally, we would like to comment on the 3D harmonic trap geometry
studied earlier in the JILA experiment \cite{Anderson2001a} and in
theoretical work by Muryshev {\it et al.}\ \cite{Muryshev1999a} and
Feder {\it et al.}\ \cite{Feder2000a}. Both experiment
\cite{Anderson2001a} and theory \cite{Feder2000a} report vortex-ring
formation during the decay of a stationary soliton (nodal-plane state)
in spherical \cite{Anderson2001a,Feder2000a} and elongated
\cite{Feder2000a} geometries at fairly high densities, which is
indicated by the nature of complex modes in the BdG spectrum of the
stationary soliton \cite{Feder2000a}. It has also been pointed out
that the BdG spectrum becomes entirely real at sufficiently low
particle number or high aspect ratio in elongated traps. However, the
decay mechanism in the presence of a single imaginary mode in the BdG
spectrum (as shown in Fig.~4 of Ref.~\cite{Feder2000a}) of a
stationary soliton in an elongated trap has not been revealed so
far. Imaginary- and real-time propagation clearly show that a
stationary svortex solution exists in this regime and that it has
lower energy than the stationary soliton. The density and phase
profile of the stationary svortex state are very similar to the
dynamically generated pattern shown in Fig.~\ref{3dharmonic}. This
figure shows the transient decay product of a perturbed stationary
black soliton after real-time propagation for 100 ms, seeded initially
with $0.01\%$ white noise. The parameters of this 3D harmonic trap
($N=10^4$ atoms of Na) correspond to Fig.~4 of Ref.~\cite{Feder2000a}
at an aspect ratio of $\omega_\rho/\omega_x = 4$ with $\omega_x
=2\pi\cdot 50\text{rad}/s$. The corresponding imaginary Bogoliubov
mode $u_j$ has an azimuthal coordinate dependence of $\exp(i\phi)$,
where $\phi$ is the azimuthal angle, and much resembles the first
imaginary BdG mode in the 2D box discussed above. The predicted decay
of the band soliton into a single svortex has not been seen, or
predicted, before and should be easily observable with current
experimental techniques.

Concluding, we have identified the fundamental modes of the SI for
transversely confined geometries: Production of 1, 2, and 3 vortices
correlates with imaginary modes in the BdG eigenvalue spectrum.
Departure from the quasi-1D regime of stability of solitons is
indicated not only by a linear stability analysis but also by the
emergence of a solitonic vortex or svortex as a stationary state of
lower energy than the corresponding dark soliton. We demonstrated
that the decay of a soliton into a single svortex is a fundamental mode of
instability in 2D box geometry and 3D elongated harmonic traps.

We like to thank Yuri Kivshar for one discussion and David Feder for
helpful discussions and the access to unpublished results. Support
from the National Science Foundation and the Alexander von Humboldt
Foundation within the Feodor Lynen program (J.~B.) is gratefully
acknowledged.


\end{document}